\documentclass[aps,twocolumn]{revtex4-1}
\usepackage{graphicx}
\usepackage[justification=justified,width=\linewidth]{caption}
\usepackage{subcaption}
\usepackage{amsmath}
\usepackage{amsfonts}
\usepackage{amsthm}
\usepackage{amssymb}
\usepackage{amsbsy}
\usepackage{wasysym}
\usepackage{bm}
\usepackage{mathrsfs}
\usepackage{color}
\usepackage{times}
\usepackage[resetlabels]{multibib}
\begin{document}
	
\title{Phase separation and rheology of segregating binary fluid under shear}
\author {Daniya Davis, Parameshwaran A and Bhaskar Sen Gupta}
\email{bhaskar.sengupta@vit.ac.in}
\affiliation{Department of Physics, School of Advanced Sciences, Vellore Institute of Technology, Vellore, Tamil Nadu - 632014, India}

\begin{abstract}
We employ molecular dynamics simulation to study the phase separation and rheological properties of a three-dimensional binary liquid mixture with hydrodynamics undergoing simple shear deformation. The impact of shear intensity on domain growth is investigated, with a focus on how shear primarily distorts the domains, leading to the formation of anisotropic structures. The structural anisotropy is quantified by evaluating domain sizes along the flow and shear direction. The rheological properties of the system is studied in terms of shear stress and excess viscosity. At low shear rates, the system behaves like a Newtonian fluid. However, the strong-shear case is marked by a transition characterized by non-Newtonian behavior.
		
\end{abstract}

\maketitle
	
\section{Introduction}

Phase separation is a fundamental process that occurs in a wide range of materials and physical systems, where a homogeneous mixture spontaneously separates into distinct phases with differing compositions~\cite{Binder,Siggia,Furukawa,Miguel,Tanaka,Beysens,Tanaka1,Kendon,Puri,Dutt,Laradji,Thakre,Ahmad}. When a homogeneous binary A-B mixture is quenched below the miscibility gap, it undergoes coarsening, leading to spinodal decomposition. This process is governed by complex thermodynamic and kinetic principles, which are further influenced by external conditions, such as temperature, pressure, and mechanical forces. Among the various external factors, shear deformation plays a critical role in determining the dynamics of phase separation in fluid mixtures~\cite{Min, Hashimoto,Onuki}. Understanding how shear influences the kinetics of phase segregation is essential for applications ranging from polymer processing and emulsions to biological systems and flow-driven material assembly.

In the absence of external forces, the phase separation kinetics of segregating fluid mixtures typically follows well-established dynamic regimes. For a binary mixture with equal concentrations of A and B (i.e., at the critical composition), an isotropic bicontinuous structure forms. This structure is characterized by a typical length scale, $\ell(t)$, representing the average domain size, that increases with time following a power law $\ell \sim t^{\alpha}$, where $\alpha$ is the growth exponent~\cite{Bray}. The value of $\alpha$ depends on the dominant transport mechanism that drives the phase separation. At early stages, domain growth is diffusion-driven, where the length scale increases as $\ell(t) \sim t^{1/3}$, following Lifshitz-Slyozov (LS) law~\cite{Chen}. As time progresses and domain size grows larger, hydrodynamic effects become significant, leading to a faster domain growth, where viscous and inertial forces begin to influence the kinetics. This transition results in new scaling laws, $\ell(t) \sim t$ for viscous hydrodynamic growth and $\ell(t) \sim t^{2/3}$ for inertial hydrodynamic regime~\cite{Siggia, Furukawa}. In general, the structure of domain evolution exhibits self-similar property, implying that the structures at two different times are same, except for the difference in size. Consequently, we observe the scaling behavior $C(r,t ) = g[r/\ell(t)]$, where $C(r,t )$ is the two-point equal time correlation function of some appropriately chosen order parameter, and $g(r)$ is the scaling function~\cite{Bray}. 

In realistic conditions and experiments, it is generally quite challenging to completely eliminate the impact of external influences on systems. Hence spinodal decomposition under external fields is highly significant in practical applications. One of the important fragment in it is the eﬀect of shear flow on spinodal decomposition. When shear is applied, the dynamics of phase separation are dramatically altered. Shear deformation induces anisotropic forces that compete with the natural coarsening of the segregating phases, leading to elongated domain structures and altered growth kinetics. In particular, the coupling between the shear flow and phase segregation results in a delicate balance between phase coarsening and the breakup of domains. This complex interplay introduces a range of new kinetic behaviors, where the growth laws deviate from those observed in unsheared systems, and domain structures often exhibit anisotropy and alignment along the shear direction. 

One of the earlier studies regarding this effect is an experiment on spinodal decomposition in a binary fluid mixture under shear flow, revealing that shear suppresses phase separation and induces anisotropic ring patterns~\cite{Perot}. The scaling behavior of these patterns aligned with the mean-field theory predictions, highlighting the significant impact of shear on phase separation dynamics. These findings underscored the importance of shear flow in controlling phase behavior in complex fluid systems. The role of hydrodynamics in such systems was investigated using light-scattering techniques \cite{chan}. Hydrodynamics was found to become important in the high shear rate regime. Evans and Morriss conducted an in-depth study of shear-thickening in fluids, highlighting the limitations of traditional profile-biased thermostats in non equilibrium molecular dynamics (NEMD) simulations~\cite{Morris}. They introduced a profile-unbiased thermostat, demonstrating its superiority in capturing high shear rate phenomena such as turbulence and string phases. Their work showed the necessity of unbiased simulation techniques for accurate modeling of complex fluid behaviors. Subsequent research focused on shear viscosity near the critical point, revealing how non-Newtonian behavior becomes pronounced due to domain deformation and breakage under shear~\cite{Onuki1}.

Padilla and Toxvaerd explored the efficient implementation of shear flow in a two dimensional fluid undergoing spinodal decomposition using the NEMD simulations~\cite{Padilla2}. They examined the use of both profile unbiased and profile biased thermostats and the achievement of steady states. The study confirmed that the Nosé-Hoover thermostat performs effectively in sheared systems with non-equilibrium flow fields. The authors also studied the phase separation of two-dimensional model binary fluid undergoing planar Couette flow \cite{Padilla3}. The anisotropic domain growth were characterized by measuring their size in different directions that followed algebraic laws. 

Later on, focusing more on the effect of shear, Corberi et al. investigated phase separation in binary mixtures under uniform shear flow using the time dependent Ginzburg-Landau free energy equation, neglecting hydrodynamics~\cite{Corberi1,Corberi2}. An accelerated domain growth was observed in the flow direction with exponent $\alpha_x = 4/3$, while the same remained unaffected in the gradient direction ($\alpha_y = 1/3$). The shear induced a log-time periodic oscillations in excess viscosity which was interpreted as stretching and breakup of domains cyclically. Attempts were made to study more of the domain structure and rheological properties using MD simulation for a two dimensional system using a profile unbiased thermostat where the shear flow was implemented by streaming velocity to the particles~\cite{Yamamoto}. In this model, the domain growth in the independent directions was minimally explored. Down the line, the effect of homogeneous shear flow on the spinodal decomposition in a two dimensional system was numerically analysed using the Cahn-Hilliard equation~\cite{Berthier}. The hydrodynamic interactions were not taken into consideration. The growth exponents were found to depend on the shear rate and different from the results obtained from the Ginzburg-Landau model~\cite{Corberi1,Corberi2}. Also, unlike in~\cite{Corberi1,Corberi2}, no oscillatory behavior was observe in the growing length scale. The dynamical steady states of the binary fluids under shear and the relative roles of diffusivity and hydrodynamics behind this were subsequently studied using the Lattice-Boltzmann method in both two and three dimensions~\cite{Cates1, Cates2}.

Despite the efforts outlined above, a complete and satisfactory understanding of the problem has not yet been fully attained due to its complexity. The role of hydrodynamics remains poorly understood at this stage. To the best of our knowledge there has been no atomistic simulation on the phase separation of liquid system under shear in three dimension with hydrodynamics taken into consideration. Also, the domain morphology analysis in all the individual directions under shear and the inter-connections between the kinetics and rheological behavior of such systems are not very well understood. Unlike the conventional experimental method where shear is applied via moving the walls which holds the system, the numerical studies introduced the shear through streaming velocity to the particles. In this paper, we undertake an extensive numerical study on the domain growth dynamics of a three dimensional immiscible symmetric binary fluid mixture in the presence of simple shear deformation. The full hydrodynamics is taken into consideration. We follow the standard experimental method where the segregating liquid mixture is placed between two parallel plates, where both the plates move laterally in opposite direction, applying shear to the system. The main aim of this work is to investigate the domain morphology and rheology of such systems for a wide range of shear rate values. 

\section{Numerical Model and Method}
The numerical model consists of binary liquid particles (A and B type) in a three dimensional simulation box, interacting via Lennard-Jones potential given as,
\begin{equation}
	U_{\alpha\beta}(r)=4\epsilon_{\alpha\beta}\left[\left(\frac{\sigma_{\alpha\beta}}{ r_{ij}}\right)^{12}- \left(\frac{\sigma_{\alpha\beta}}{r_{ij}}\right)^6\right]
\end{equation}
where $\epsilon$ is the interaction strength, $\sigma$ is the particle diameter, $r=|r_i-r_j|$ is the scalar distance between the two particles $i,j$ and $\alpha,\beta \in A,B$. The phase separation between the two species occurs when at low temperatures, the positive excess energy of mixing exceeds the entropy of mixing. This is assured in the simulation by assigning the inter-particle diameters as $\sigma_{AA}=\sigma_{BB}=\sigma_{AB}=1.0$ and the interaction parameters as $\epsilon_{AA}=\epsilon_{BB}=2\epsilon_{AB}=1$. The computational load is reduced by assigning  a cutoff at $r=r_c=2.5$ to the LJ potential.

To encounter the discontinuity in the potential and force terms caused by the introduction of cut-off is resolved by modifying the potential as, 
\begin{equation}
	\label{eq:mod_potential}
	u(r)=U(r)-U(r_c)-(r-r_c)\left(\frac{dU}{dr}\right)|_{r=r_c} 
\end{equation}
The final term of the modified potential takes care of the discontinuity problems. The critical temperature and critical density of the above mentioned system is $T_c=1.42$ and $\rho_c=\frac{N}{V}=1.0$ in three dimension \cite{DasTc}, N and V being the number of particles and volume of the system. The temperature and length are measured in reduced LJ units of $\frac{\epsilon}{k_B}$ and $\sigma$ respectively. For simplicity, we consider $\epsilon,\sigma$ and $m$ equal to unity.

We consider an elongated box with dimensions $L_x \times L_y \times L_z = 192 \times 64 \times 64$, incorporating two walls located in the $xy$-plane at $z = 0$ and $z = L_z$. These walls are constructed from closely packed particles (denoted as $P$-type), similar to the fluid particles. The binary fluid interacts with the walls through standard cross-species interactions, where $\epsilon_{AP} = \epsilon_{BP} = 0.5$. Shear is introduced by moving the walls laterally in opposite directions: the wall at $z = 0$ moves with a constant velocity $v_x$ in the negative $x$-direction, while the wall at $z = L_z$ moves with the same velocity in the positive $x$-direction. This generates an effective shear rate $\dot{\gamma} = 2v_x/L_z$. Periodic boundary conditions are applied in the $x$ and $y$ directions.

To study phase separation, we employ molecular dynamics (MD) simulations in the canonical ensemble. Since the system is in the liquid state, it is crucial to account for hydrodynamic effects. Therefore, we use the Nos\`{e}-Hoover thermostat, which maintains the desired temperature while preserving the hydrodynamics~\cite{Nose}. The velocity-Verlet algorithm is employed to update the positions and velocities of the particles~\cite{Verlet}, with a time step of $\Delta t = 0.005$, where time is measured in LJ units of $(\frac{m\sigma^2}{\epsilon})^{\frac{1}{2}}$. The binary liquid system consists of $N=629144$ particles, with equal numbers of the two species, and an average density of $\rho = 0.8$.

The simulation begins by equilibrating the system at a high temperature of $T_i = 6.0$, producing a homogeneous mixture in the absence of shear. At time $t = 0$, the system is suddenly quenched to $T_f = 1.0$, well below the bulk critical temperature $T_c$. Shear is applied at a constant rate immediately after the quench, and the system is allowed to evolve toward the thermodynamically favored state. We analyze phase separation under a wide range of shear rates, specifically $\dot{\gamma} = 0.001, 0.003, 0.01, 0.02, 0.03, 0.06$. This range ensures that the system remains in the linear response regime, where the homogeneous thermostat functions effectively. Thus, phase separation occurs under shear without introducing nonlinear effects that could complicate the  interpretation of results. Ensemble averages of all statistical quantities are obtained from 20 independent runs, each starting from a completely different initial configuration.

\section{Results}
\subsection{Domain morphology under shear}
\begin{figure}[h]
\centering
    \includegraphics[width=1.0\columnwidth]{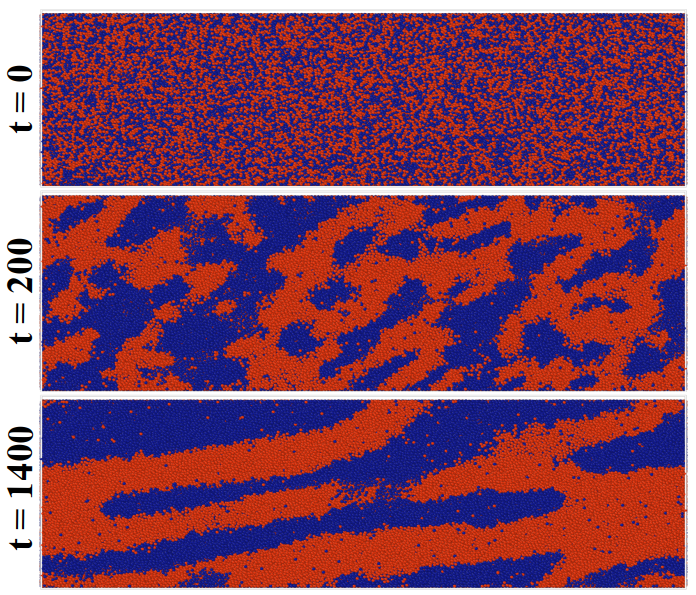}
    \caption{Time evolution snapshots of the binary phase separating system under shear with $\dot{\gamma}=0.01$. The two dimensional cross sectional area on the $xz$ plane is shown. Red and Blue colors signify A and B types of particles respectively.}
     \label{fig:fig.01}
\end{figure}

Fig.~\ref{fig:fig.01} presents the evolution snapshots of the phase separation dynamics for our segregating liquid mixture system under an applied shear rate of $\dot{\gamma} = 0.01$. In the absence of external forces, such a system would typically separate into isotropic domains, minimizing the surface area and interfacial energy between the two phases. However, the application of shear through the movement of the top and bottom walls introduces an anisotropic field, causing the liquid layers to slide past each other. As a result, the domains elongate along the shear direction. Fig.~\ref{fig:fig.01} clearly shows that the domains not only evolve in size, as seen in standard phase separation, but also stretch along the flow direction ($x$-axis). This demonstrates that the shearing effect (directional dependence) in the liquid can be accurately captured in atomistic simulations by employing the conventional experimental method of moving walls.
\begin{figure}[b]
\centering
    \includegraphics[width=0.5\textwidth]{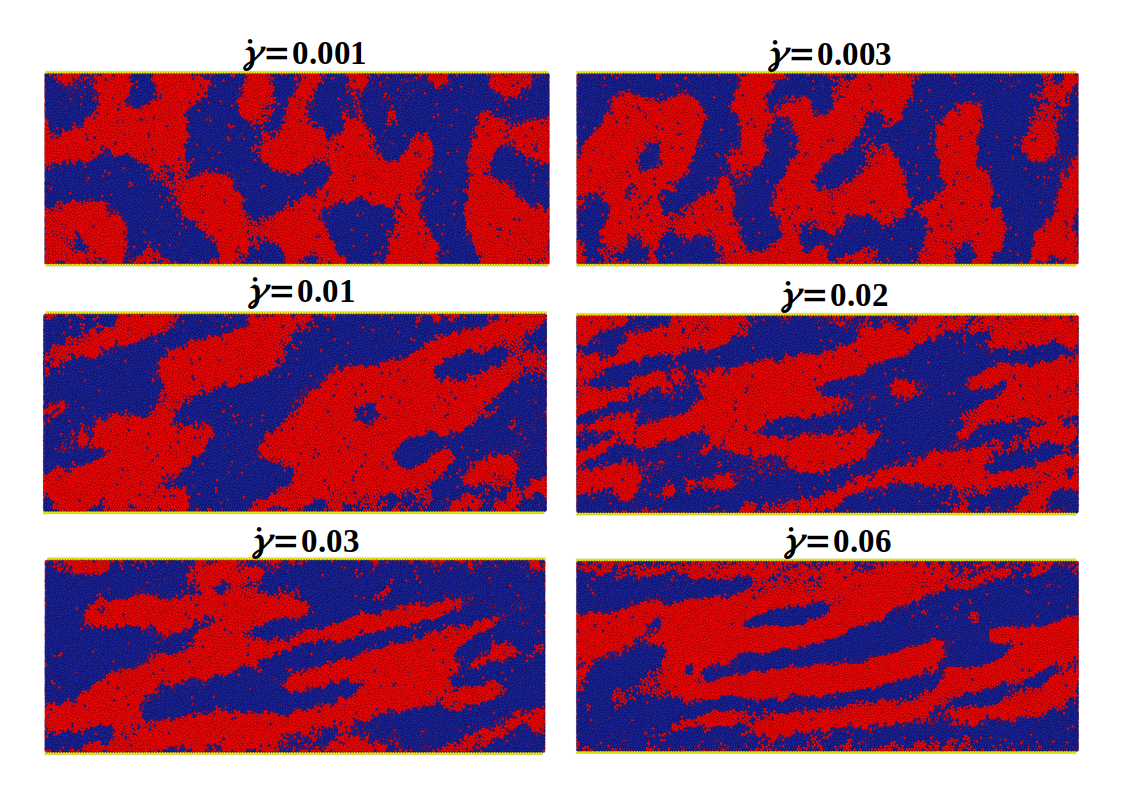}
    \caption{Time evolution snapshots of the binary phase separating system for different shear rates at a fixed time $t=500$. The two dimensional cross sectional area on the $xz$ plane is shown. Red and Blue colors signify A and B types of particles respectively.}
    \label{fig:02_snapshots_500t}
\end{figure}

To analyze the dependence of domain morphology on the shear strength, we display in fig.~\ref{fig:02_snapshots_500t} the representative snapshots of the segregating system subjected to different shear rates at time $t=500$. At lower shear rates ($\dot{\gamma} =0.001,0.003$), the domains appear isotropic (nearly spherical and less elongated) and closely resemble to those of the unperturbed system. However, as the shear rate is increased to \(\dot{\gamma} = 0.01\), the domains start to elongate along the flow direction. With further increase, the domains become increasingly stretched resulting in highly anisotropic structures. This phenomena indicates that the shear force dominates over the domains' natural tendency to minimize interfacial energy. The dependence of domain morphology on shear strength underscores the complex interplay between external forces and phase separation dynamics. 

To characterize the domain morphology and study the growth dynamics, we resort to the two-point equal time correlation function $C(r,t)$~\cite{Rounak1,Rounak2, Rounak3}. As shown in figs.~\ref{fig:fig.01} and~\ref{fig:02_snapshots_500t}, the domains do not evolve isotropically. Therefore, we compute the correlation separately in each direction given as,
\begin{equation}
	\label{eq:correlation}
	C(r_e,t)=\langle\psi(0,t)\psi(r_e,t)\rangle - \langle\psi(0,t)\rangle\langle\psi(r_e,t)\rangle 
\end{equation} 
where $e=\in x,y,z$ and $r_e=e\hat{e}$. $\hat{e}$ is the unit vector~\cite{Davis1, Davis2}. The angular brackets represent ensemble averaging. $\psi(r,t)$ is the local order parameter computed from the local density difference between species A and B within a box of size $(2\sigma)^3$ centered at position $\vec{r}$. The value of $\psi(r,t)$  is set to +1 if species $A$ dominates at that site, and -1 otherwise. 

The Fourier transform of the correlation function gives the structure factor given by,
\begin{equation}
	\label{eq:structure_factor} 
	S(\vec k,t)= \int d\vec r\ exp(i\vec k.\vec r)\ C(\vec r,t) 
\end{equation}
Due to the anisotropic domain growth under shear, the structure factor is calculated separately for each spatial direction instead of using spherical averaging~\cite{Paramesh}. 

\begin{figure}[t!]
\centering
    \includegraphics[width=1.0\columnwidth]{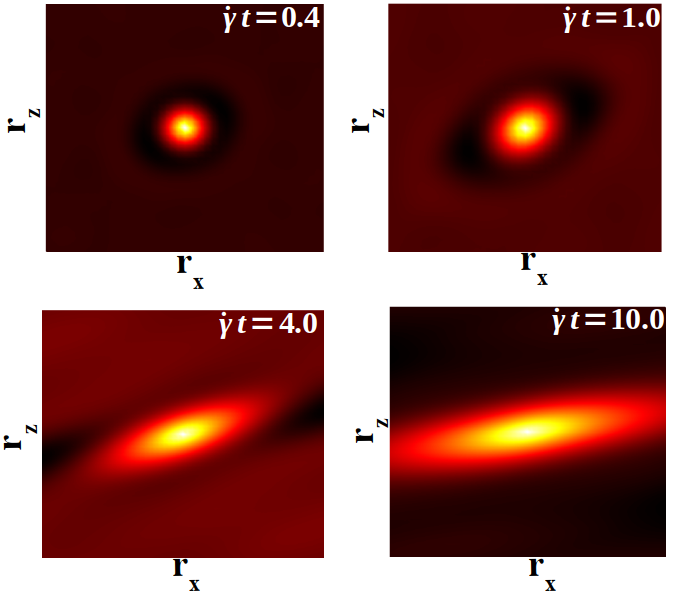}
    \caption{The time evolution of the two-point equal time correlation function is shown in the $xz$ plane (see text).}
    \label{fig:03_correlation_vis}
\end{figure}
\begin{figure}[b!]
\centering
   \includegraphics[width=1.0\columnwidth]{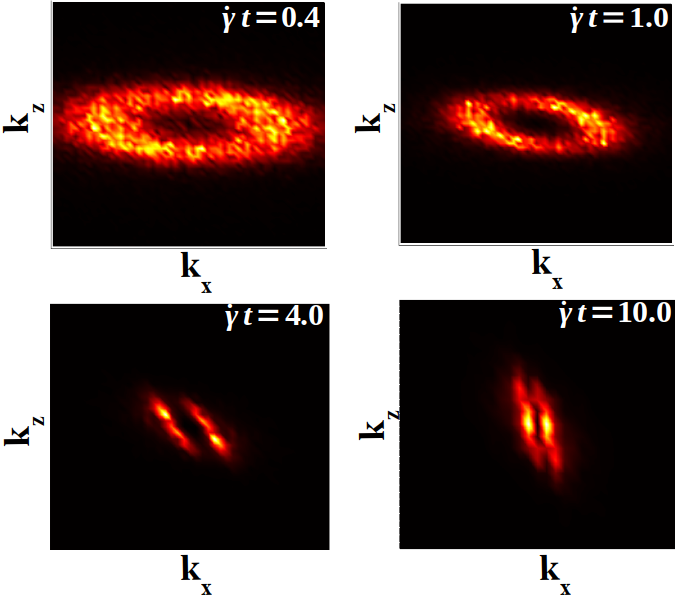}
    \caption{The time evolution of the structure factor is shown in the $xz$ plane (see text).}
    \label{fig:04_scattering}
\end{figure}

To understand the anisotropic domain growth, we analyze the correlation function given by Eq.~\ref{eq:correlation}.  Fig.~\ref{fig:03_correlation_vis} shows a visual representation of the spatial correlation for $\dot{\gamma}=0.01$. Initially, the domains are small and spherical, indicating that the system is just beginning to separate into distinct phases with minimal influence of the shear. As time progresses, the correlation function reveals that the domains start to elongate and stretch along the flow direction. This elongation indicates anisotropic domain growth. At large deformation, the correlation collapses entirely along the x-axis, signifying that the domains have become highly elongated and predominantly aligned with the flow direction. The analysis of the spatial properties is extended by studying the corresponding scattering patterns $S(\vec k,t)$ as shown in Figure~\ref{fig:04_scattering}. This offers a complementary perspective to the correlation function by illustrating how the domains scatter incident waves, further highlighting the anisotropy in the system. 

\begin{figure}[b]
\centering
	{\includegraphics[width=\columnwidth]{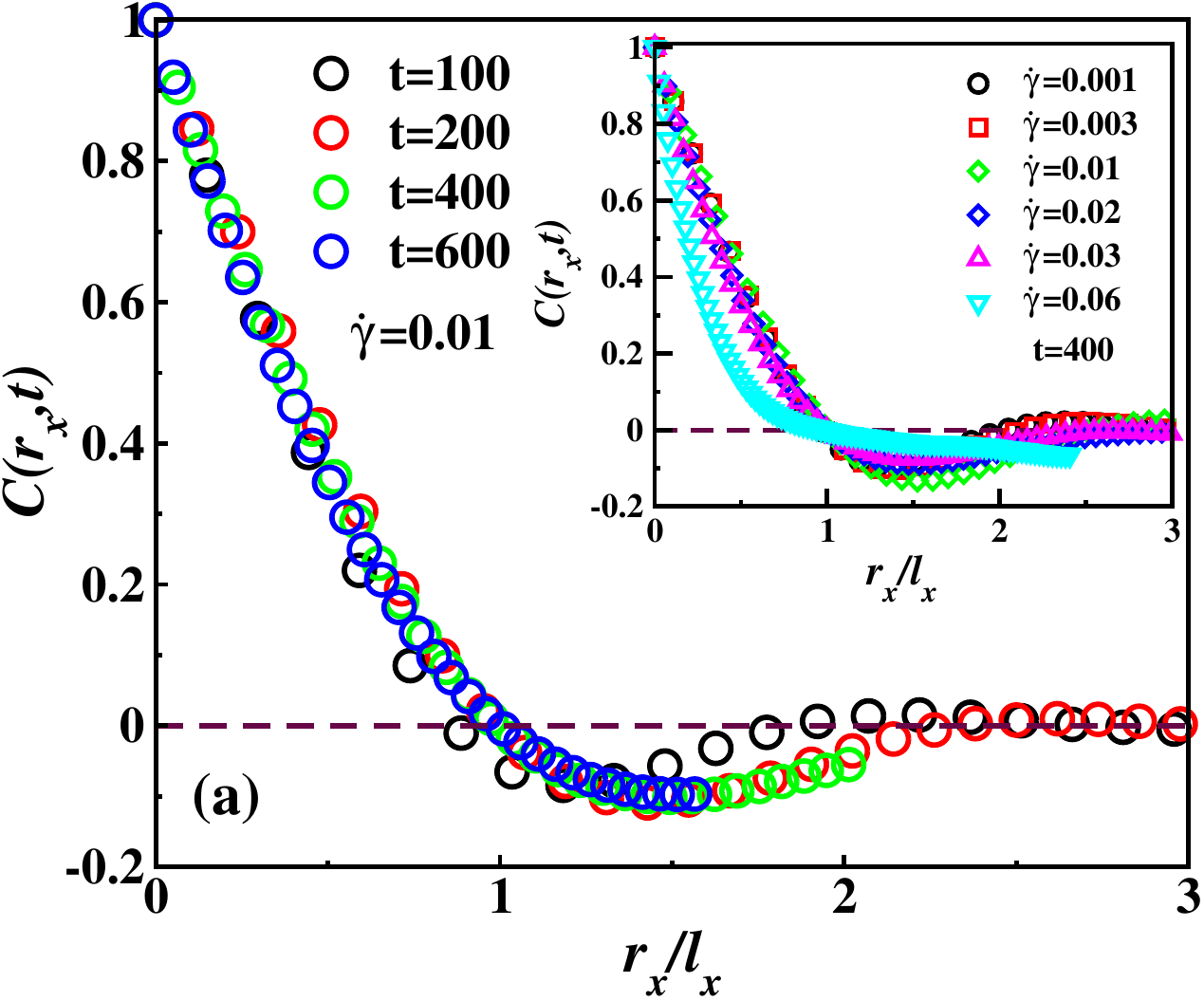}}
	\hfill
    {\includegraphics[width=\columnwidth]{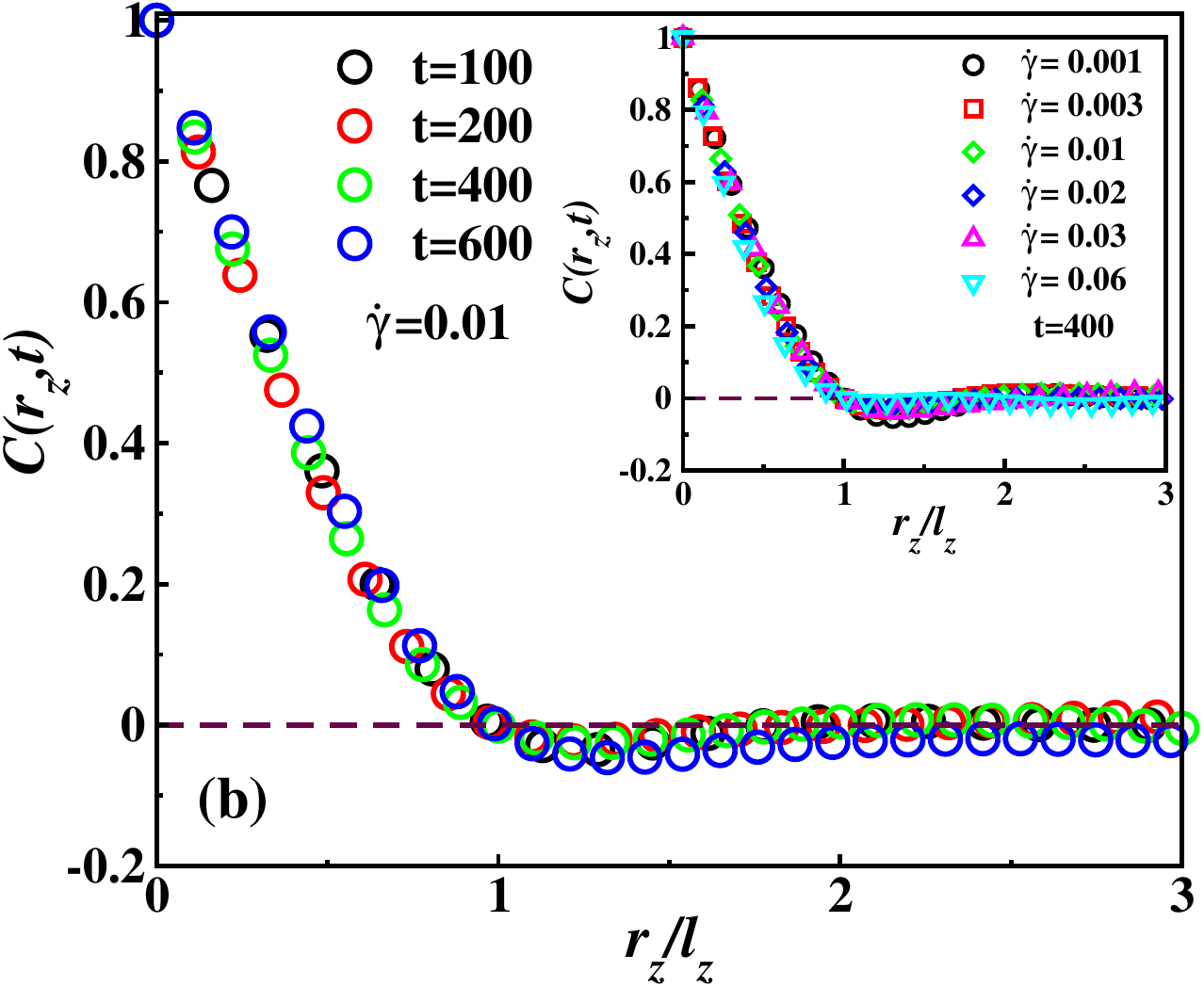}}
    \caption{In (a) we show the scaling plot of the correlation function $C(r_x,t)$ vs $r_x/\ell_x$ at $\dot{\gamma} = 0.01$ for different time in the flow direction. The inset shows the scaling plot for different $\dot{\gamma}$ values at a fixed time $t=400$. In (b) the same scaling plots in the gradient direction is shown. }
    \label{fig:05_rx_rz_scaled}
\end{figure}

To quantitatively understand the domain growth, we examine the scaling behavior of the correlation function in the flow and gradient directions. Fig.~\ref{fig:05_rx_rz_scaled}(a) presents the scaled $C(r_x,t)$ vs $r_x/\ell_x$ plot for different times at a shear rate of $\dot{\gamma} =0.01$ in the flow direction. The average domain size, 
$\ell_x$, is defined as the first zero crossing of $C(r_x,t)$. The overlap of the curves indicates that the domain structures maintain a self-similar nature over time, meaning that while the domains grow, their overall shape and distribution remain consistent. The slight deviation in overlap at the initial time ($t=100$) corresponds to the transition phase, where the domains shift from an isotropic shape to an elongated one. This brief period of non-overlap reflects the morphological changes as the system adapts to the shear flow. Note that, self-similar domain growth is observed for all the chosen strain rate values. Due to the repetitive nature of the data, these additional cases are not shown here.

The inset of fig.~\ref{fig:05_rx_rz_scaled}(a) shows the scaled $C(r_x,t)$ vs $r_x/\ell_x$ curves for all shear rates at a fixed time $t=400$. Barring for $\dot{\gamma} =0.06$, a satisfactory data collapse is observed. However, for $\dot{\gamma} =0.06$, the overlap is significantly poor, indicating that the domain structures differ considerably from the rest. This suggests that except for very high shear rates, the domains morphology follows the same universality class. 

In fig.~\ref{fig:05_rx_rz_scaled}(b) we show the scaled correlation functions in the gradient direction for the shear rate $\dot{\gamma} =0.01$. The same scaling plot for different strain values is shown in the inset at a fixed time $t=400$. Both the cases exhibit excellent data collapse. It is important to note that, since the shear effect is negligible in the $y$-direction, we abstain from showing those. These results convincingly demonstrate the universality and self-similar nature of domain growth in all directions.

\begin{figure}[h]
\centering
	{\includegraphics[width=\columnwidth]{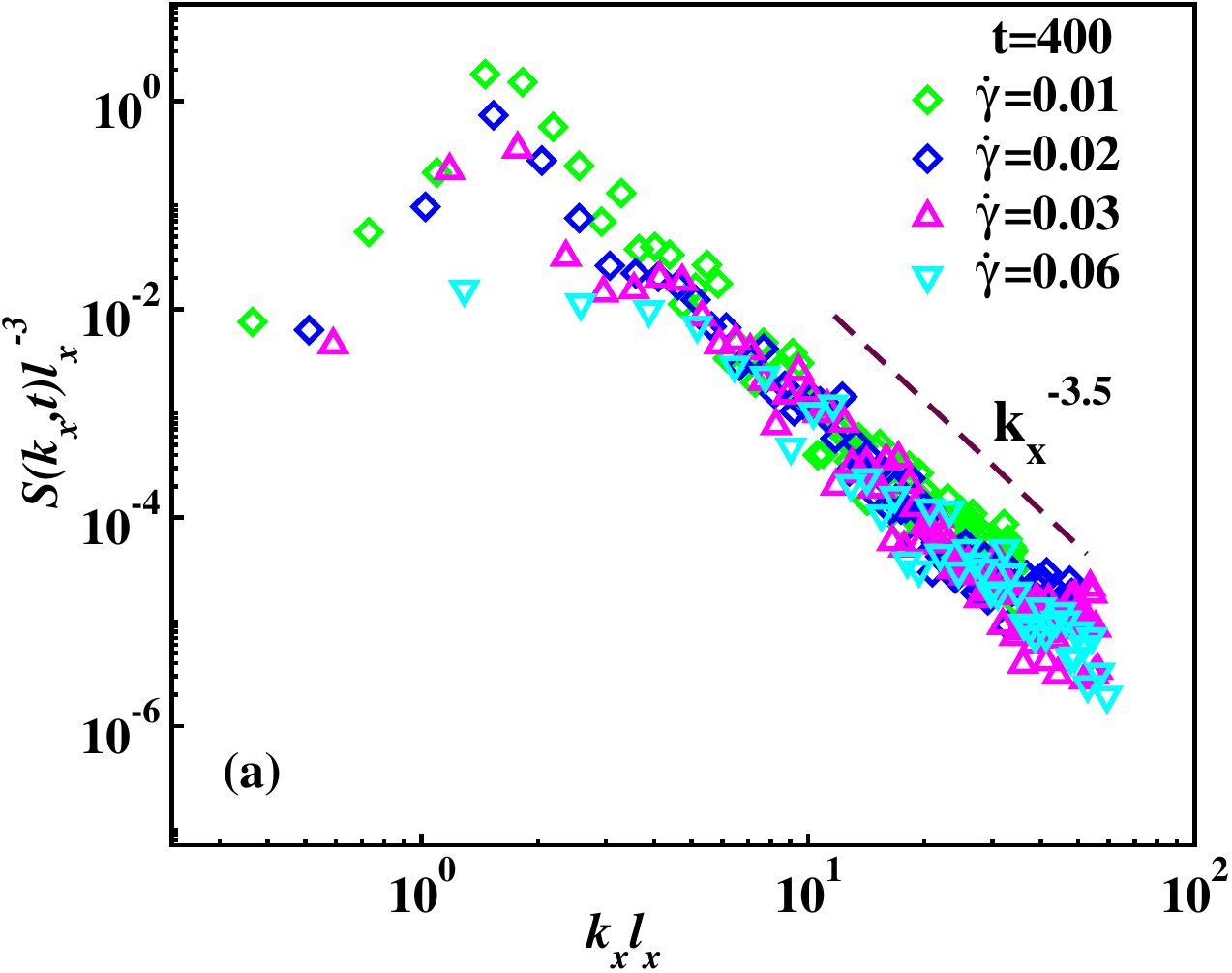}}
	\hfill
    {\includegraphics[width=\columnwidth]{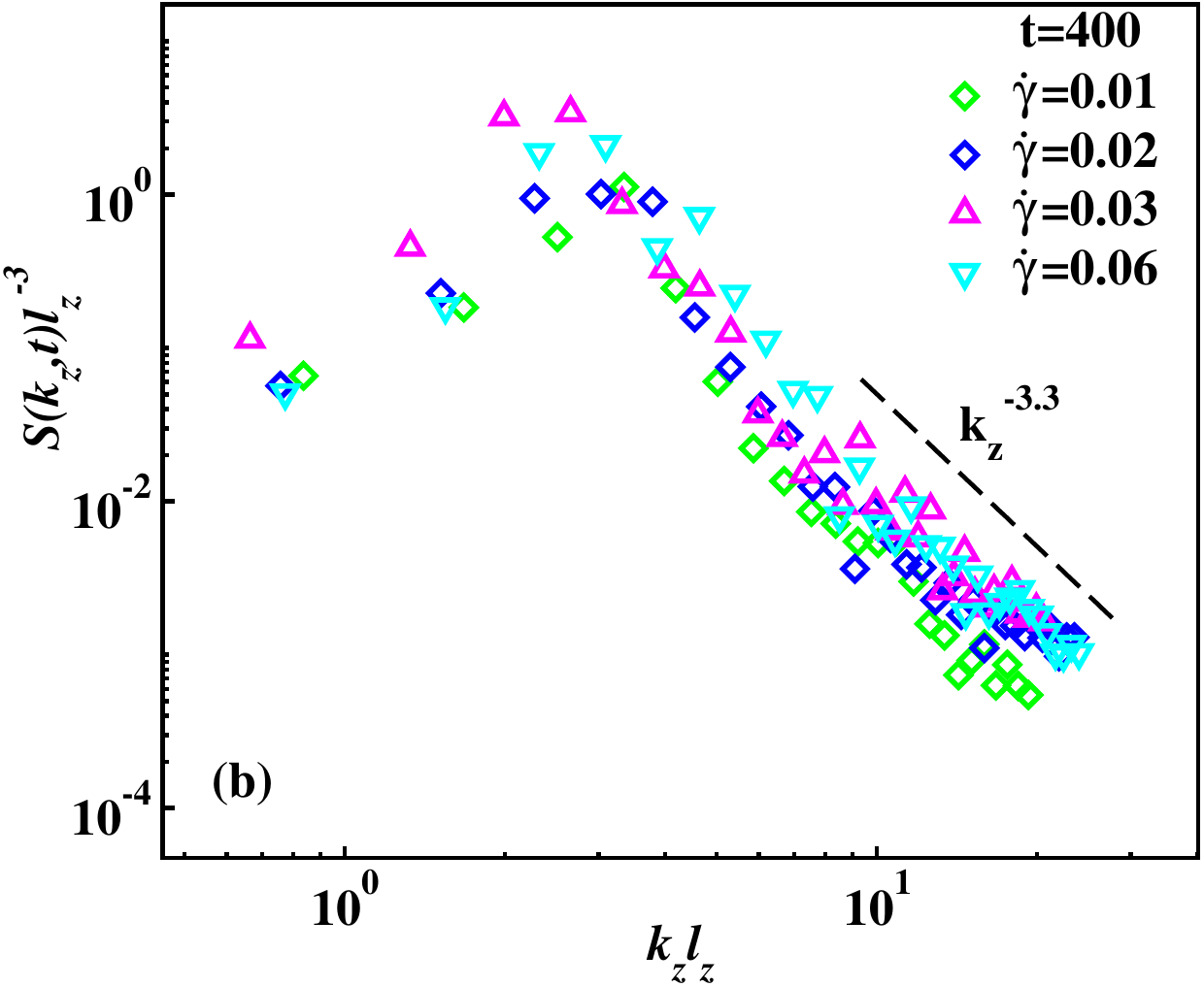}}
    \caption{The scaled structure factor for different $\dot{\gamma}$ values at a fixed time $t=400$ in the flow and gradient directions are shown in (a) and (b) respectively. The dashed lines are the reference for the power law behavior.}
    \label{fig:06_kx_kz_scaled}
\end{figure}

Fig.~\ref{fig:06_kx_kz_scaled}(a) presents the scaled structure factor $S(k_x, t)\ell_x^{-3}$ versus $k_x\ell_x$ in the flow direction at $t = 400$. The tail part of $S(k_x, t)$ for the of the unperturbed system follows the Porod law behavior $S(k_x, t) \sim k_x^{-(d+1)}$ \cite{Bray}. Since the domains at low shear rate resembles the zero shear case, we show results for $\dot{\gamma} \geq 0.01$. We observe the higher shear rate modifies the large $k$ behavior and a non-Porod tail $S(k_x, t) \sim k_x^{-(d+\theta)}$ is observed. This can be attributed to the roughening of the domain interfaces with fractal dimension $d_f=d-\theta$ due to shear \cite{gaurav}. The roughness is found to increase with shear rate. For the largest shear rate $\dot{\gamma} =0.06$, we find $\theta=0.5$ and $d_f=2.5$. We observe an enhancement in the roughening of the domain boundary in the gradient direction as shown in fig.~\ref{fig:06_kx_kz_scaled}(b). The corresponding fractal dimension is found to be $d_f=2.7$.

\subsection{Domain growth dynamics}
\begin{figure}[t!]
\centering
   {\includegraphics[width=1.0\columnwidth]{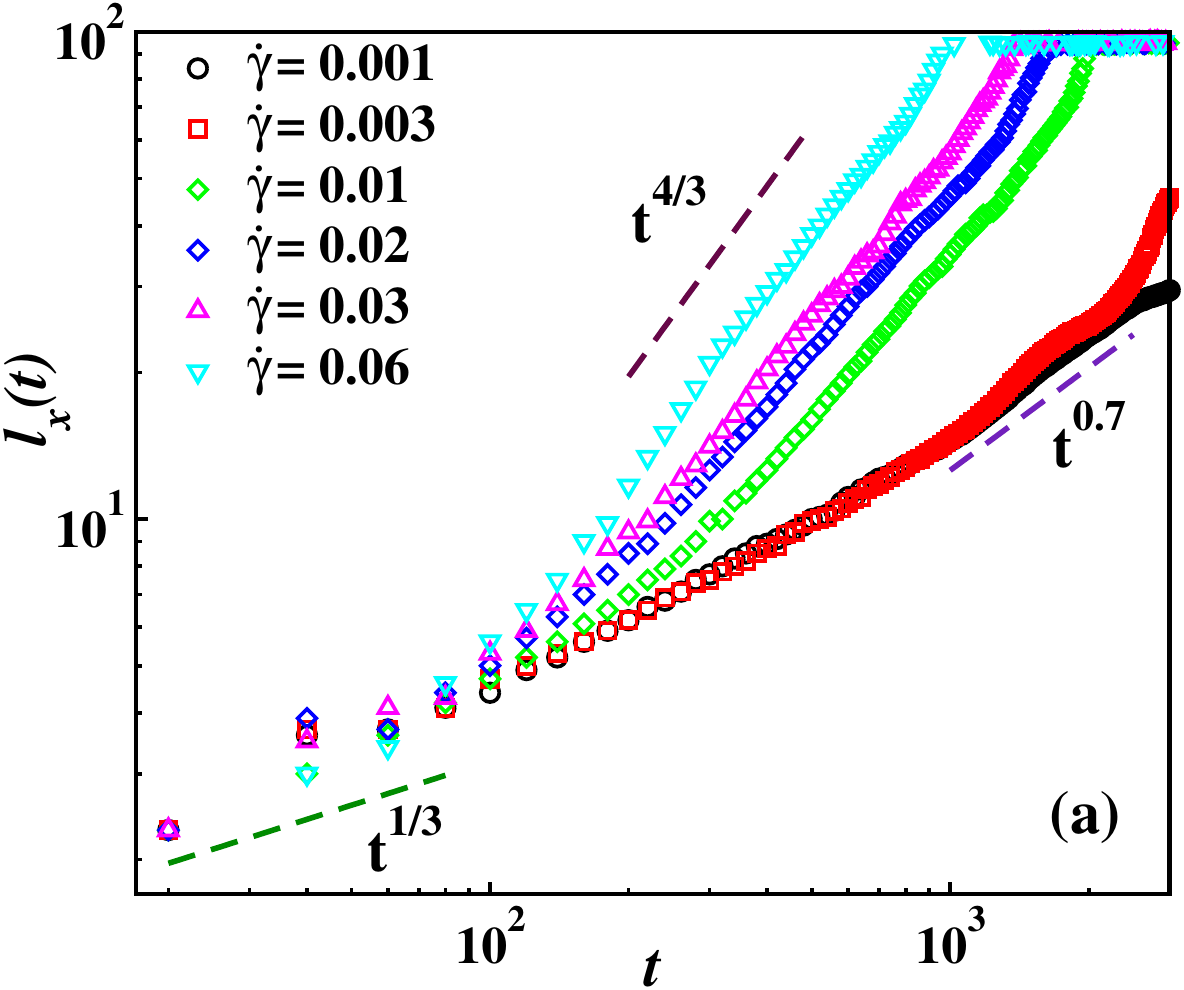}}
   {\includegraphics[width=1.0\columnwidth]{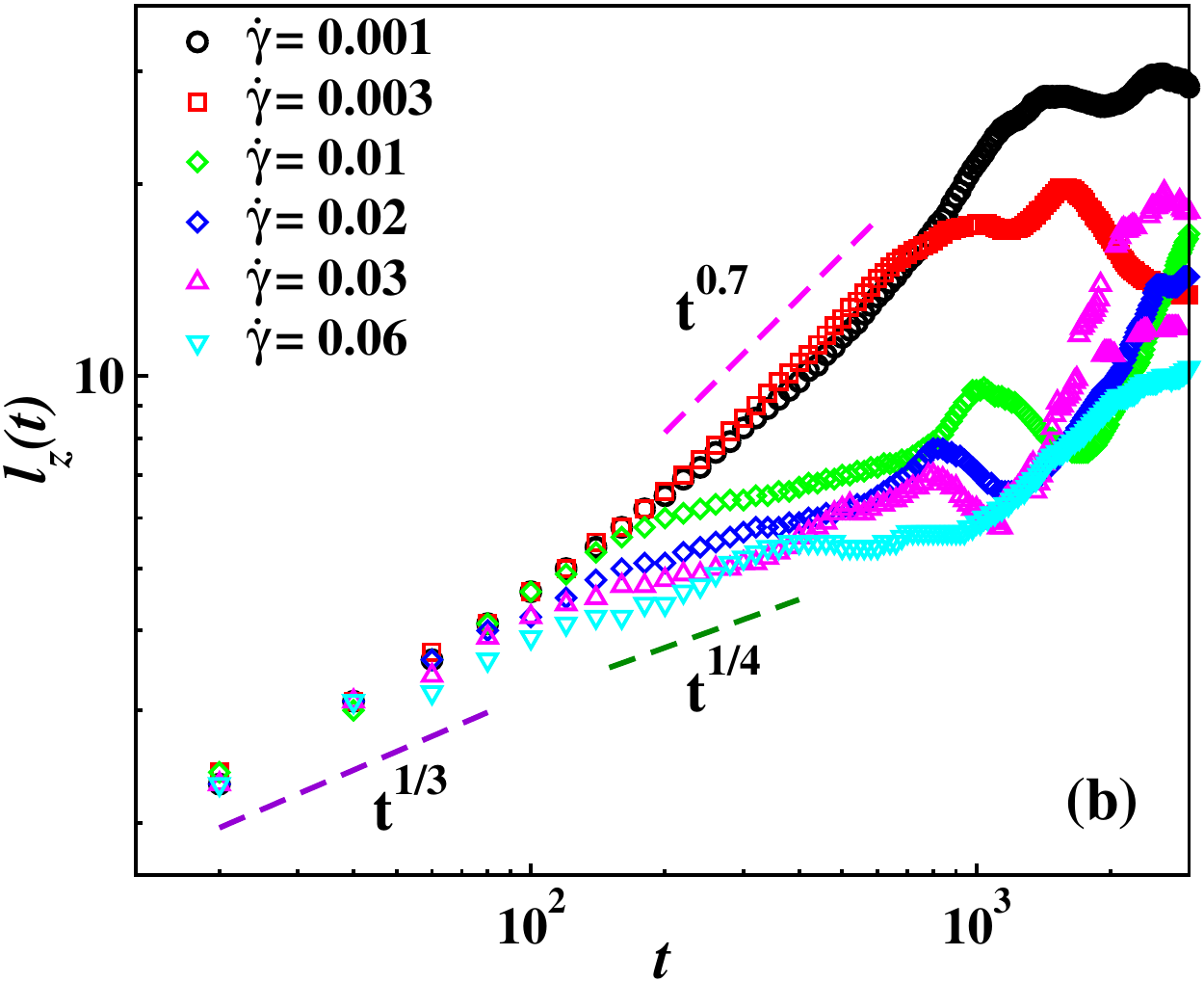}}
    \caption{The time evolution of the length scale for different $\dot{\gamma}$ values in the flow and gradient directions are shown in (a) and (b) respectively. The dashed lines are the reference for the power law growth.}
    \label{fig:07_x_lengthscale}
\end{figure}

In this section, we quantify domain growth in terms of the length scale $\ell(t)$. We start our analysis in the flow direction.  In a hydrodynamics preserving three-dimensional fluid system, the growth exponent is $\alpha = 1.0$ after an initial short-time diffusive regime referred to as Lifshitz-Slyozov law ($\alpha=1/3$)~\cite{Chen}. However, in the presence of moving walls, the domain growth is expected to differ from that of an unperturbed system. Fig.~\ref{fig:07_x_lengthscale}(a) shows the time evolution of the length scale $\ell_x(t)$. Our simulations accurately capture the initial diffusive regime with growth exponent $\alpha=1/3$. For low shear rates ($\dot{\gamma} < 0.01$), the  growth remains similar to that of the unperturbed system. In fig.~\ref{fig:07_x_lengthscale}(a), for $\dot{\gamma} = 0.001$ and  0.003, the exponent $\alpha \sim 0.7$ corresponds to viscous hydrodynamic growth. The slight deviation from unity is attributed to the nonzero offset during the crossover~\cite{Majumdar}. Therefore, at lower shear rates, the growth dynamics in the flow direction are similar to those in the absence of shear.

As the shear rate is increased ($\dot{\gamma} \geq 0.01$), a new length scale emerges after the diffusive regime, representing a faster domain growth. This crossover occurs earlier for systems with higher $\dot{\gamma}$ values. The power-law growth with an exponent $\alpha = 4/3$ is represented by the dashed line in the figure. This long-time growth is consistent across different high shear rates, suggesting that once a certain threshold is crossed, the shear uniformly enhances the domain growth in the flow direction. A similar behavior of the growth exponent $\alpha$ that remains unaffected in the presence of shear is observed in experiment \cite{Lauger}.

Next, we turn our focus to the velocity gradient direction. The results are shown in fig.~\ref{fig:07_x_lengthscale}(b). Significant fluctuations are observed in the domain growth pattern, depending on the strength of the applied shear field. At short times, the growth follows the Lifshitz-Slyozov (LS) law with an exponent of $\alpha = 1/3$. In the hydrodynamic regime, for lower shear rates ($\dot{\gamma} < 0.01$), the exponent is approximately $\alpha \approx 0.7$, indicating that the growth in unaffected by shear. In contrast, for higher shear rates ($\dot{\gamma} \geq 0.01$), we observe a significant change in $\alpha$. The growth rate in this regime corresponds to $\alpha \approx 1/4$. This sharp decline demonstrates that high shear strongly suppresses domain growth in the velocity gradient direction, resulting in a highly anisotropic growth pattern. The same behavior is observed in real experiments \cite{chan}.
\begin{figure}[h!]
\centering
   {\includegraphics[width=1.0\columnwidth]{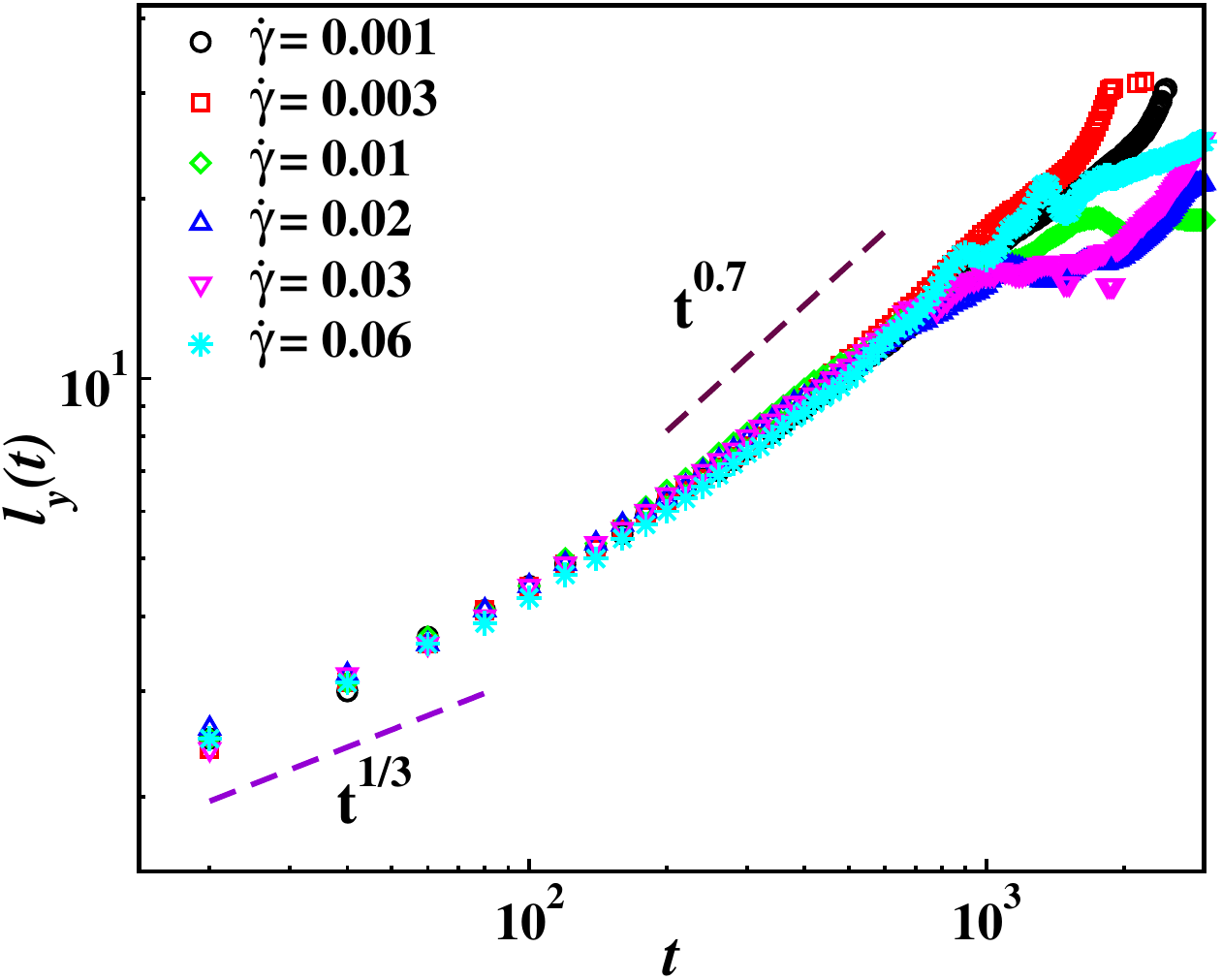}}
    \caption{The time evolution of the length scale for different $\dot{\gamma}$ values in the vorticity direction. The dashed lines are the reference for the power law growth.}
    \label{fig:09_y_lengthscale}
\end{figure}

According to the analytical results from the renormalization group theory, the relation between growth exponents in the flow and gradient directions is given as $\Delta\alpha=\alpha_x-\alpha_z=1$ \cite{Corberi3,Rapapa,Bray1}. The experimental results measured the $\Delta\alpha$ in the range 0.8-1.0 \cite{Lauger,chan1}. From our simulation we find $\Delta\alpha=4/3-1/4=1.09$, which matches pretty well with the theoretical predictions.  

Next, we examine the growth dynamics in the vorticity direction ($y$-direction), where neither flow nor velocity gradient is present. In this direction, after the initial diffusive growth with an exponent of $\alpha = 1/3$, the system enters the viscous hydrodynamic regime with a growth exponent $\alpha \approx 0.7$. Within the linear response regime of the shear rate, the growth rate remains independent of $\dot{\gamma}$. Therefore, the domain growth in the vorticity direction remains unaffected by the shear.

\begin{figure}[b!]
\centering
   {\includegraphics[width=1.0\columnwidth]{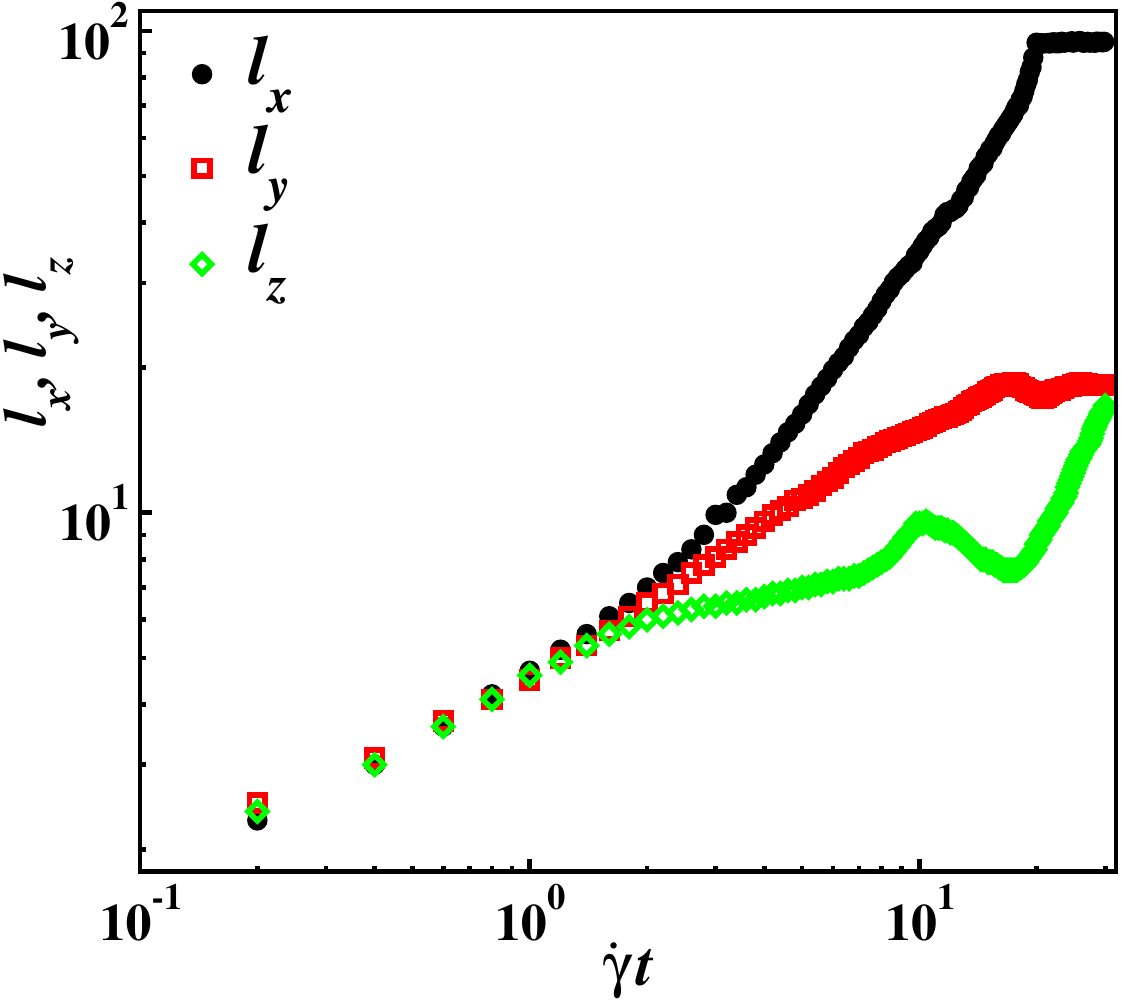}}
    \caption{The time evolution of the length scale for a fixed $\dot{\gamma}=0.01$ in all the directions.}
    \label{fig:10_lengthscale_0.01}
\end{figure}
Our results clearly indicate that domain growth is uncorrelated across all directions. For clarity, we present the time evolution of the length scale in Fig.~\ref{fig:10_lengthscale_0.01} for a fixed $\dot{\gamma} = 0.01$. To the best of our knowledge, this is the first atomistic study to analyze the domain growth of a binary fluid with hydrodynamics in three dimension under shear deformation, mimicking standard experimental conditions.

\subsection{Rheology during phase separation}
Finally, we focus on the rheological properties of the system during phase separation under shear. We begin our analysis by computing the stress tensor for the system, defined as:
\begin{equation}
	P_{ij} = \frac{1}{V} \sum_{k=1}^{N} m_k v_{k_{i}} v_{k_{j}}  + \frac{1}{V} \sum_{k=1}^{N`} r_{k_{i}} f_{k_{j}}
\end{equation}
where  $i,j \in {x,y,z}$, $m_k = 1$ is the mass of the individual particle, $r_{k},  v_{k}, \and f_{k}$ are the position, velocity and force components respectively~\cite{Yamamoto}. 

\begin{figure}[h]
\centering
   {\includegraphics[width=1.0\columnwidth]{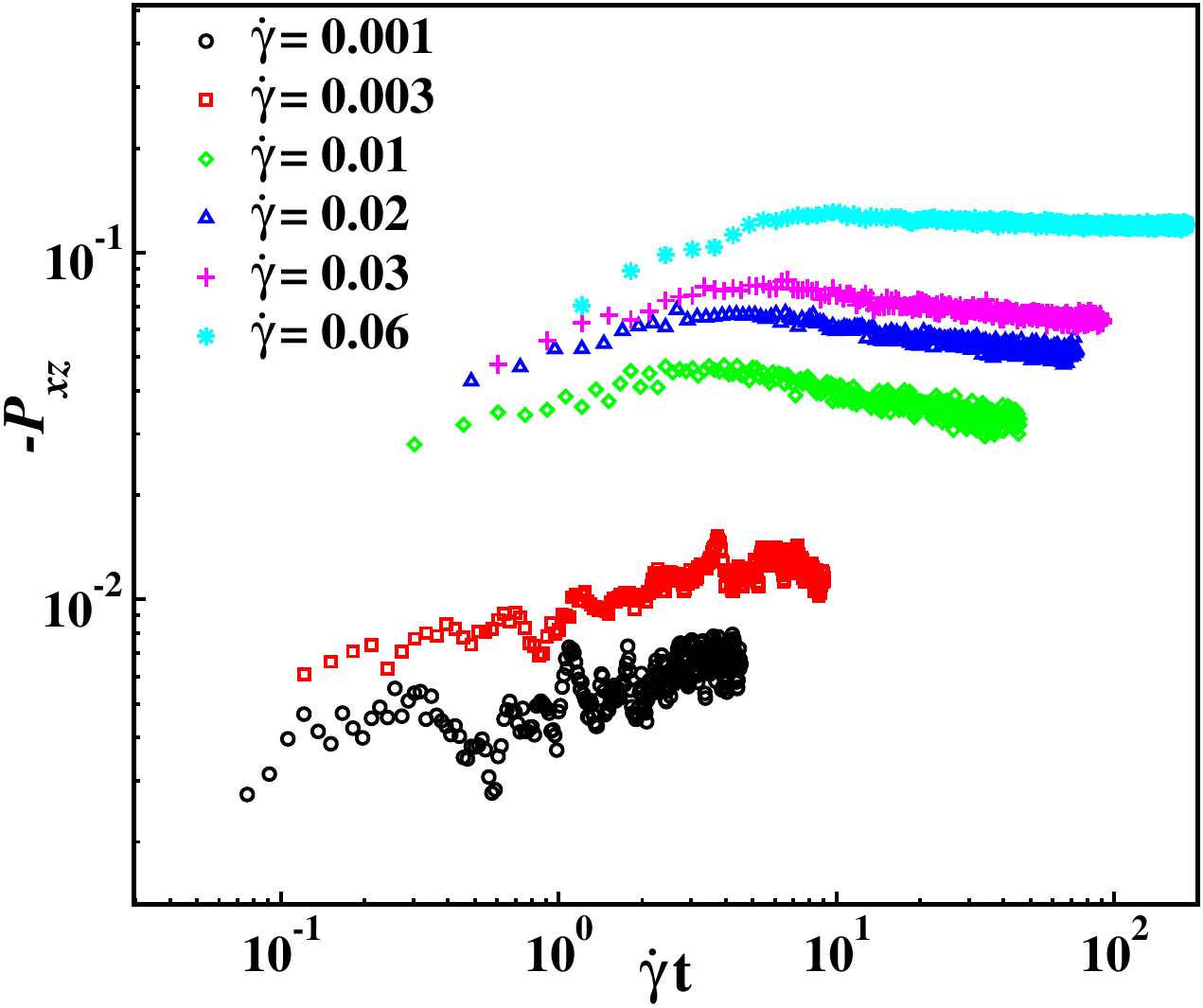}}
    \caption{The plot of shear stress $-P_{xz}$ vs shear strain $\dot{\gamma}t$ for different $\dot{\gamma}$ values mentioned in the inset.}
    \label{fig:11_stress_strain}
\end{figure}
We plot the temporal shear stress defined as $-P_{xz}$ against the shear strain $\gamma = \dot{\gamma} t$ in Fig.~\ref{fig:11_stress_strain}. For lower shear rates ($\dot{\gamma} < 0.01$), the plots show significant noise, which can be attributed to the competition between the shear field and the domain coarsening process. While domains are forming and evolving, the applied shear simultaneously tries to deform these domains, creating turbulence and resulting in fluctuations in the stress data.

For higher shear rates ($\dot{\gamma} \geq 0.01$), a systematic increase in stress with strain is observed, followed by a decrease and then stabilization at a lower value. Initially, as domains form, they resist the high shear field, causing an increase in stress. This demonstrates the viscoelastic nature of the fluid. After this initial rise, the stress decreases as the domains elongate along the shear direction and break periodically, reducing the resistance. Eventually, the system reaches a steady state, and the stress stabilizes at a lower value.
\begin{figure}[h]
\centering
   {\includegraphics[width=1.0\columnwidth]{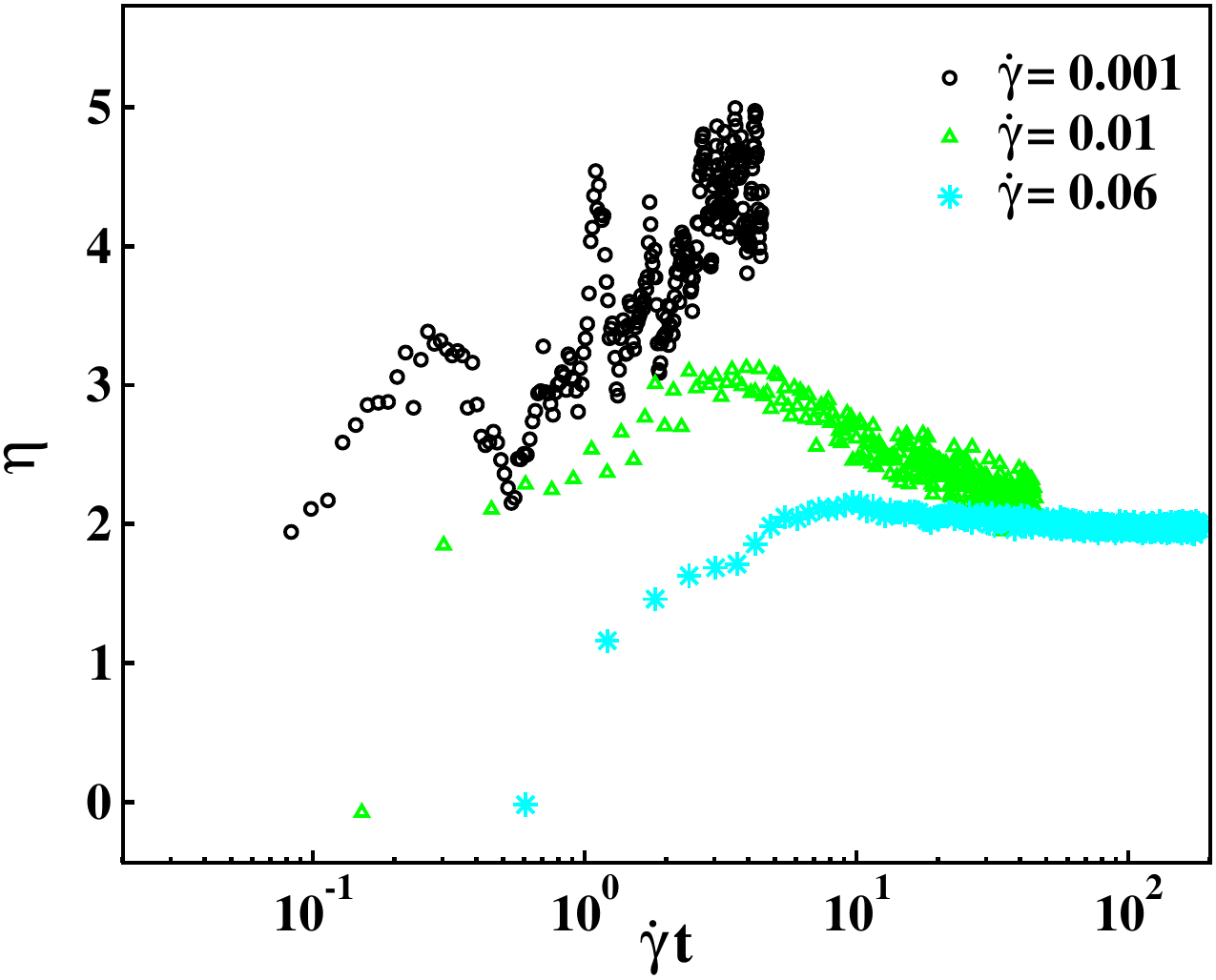}}
    \caption{The shear viscosity $\eta$ vs shear strain $\dot{\gamma} t$ plot for three different shear rates mentioned in the inset.}
    \label{fig:12_visc_strain}
\end{figure}

Next we focus on one of the most important rheological property of the system, the shear viscosity, which quantifies the flow under shear. It measures how easily a fluid can be shaped or deformed by an external force~\cite{Padilla3}. The shear viscosity is calculated as follows,
\begin{equation}
	\eta = \frac{-P_{xz}}{\dot{\gamma}}
\end{equation}
Fig.~\ref{fig:12_visc_strain} shows the shear viscosity vs shear strain for three different shear rates $\dot{\gamma}=0.001, 0.01$ and 0.06. The curve for $\dot{\gamma} =0.001$ is quite noisy due to the system's low response at this shear rate. Even so, the instantaneous value of the shear viscosity for a given strain value is higher for $\dot{\gamma} =0.001$ compared to the other strain values.

In contrast, the curves for the other two shear rates are much smoother. For $\dot{\gamma} = 0.01$ the viscosity initially increases to a peak and eventually, as the domains deform, the viscosity decreasing to a steady value. For highest shear rate of $\dot{\gamma} =0.06$, the change in viscosity is less pronounced, showing only a small initial increase that quickly stabilizes. In the high-shear regime, the viscosity values converge closely in the large deformation limit, highlighting a significant difference between the behavior in weak and strong shear regimes. A similar behavior is observed in experiments where the steady state value of the viscosity is the same for different shear rates \cite{Lauger}. 
\begin{figure}[bh!]
\centering
   {\includegraphics[width=0.99\columnwidth]{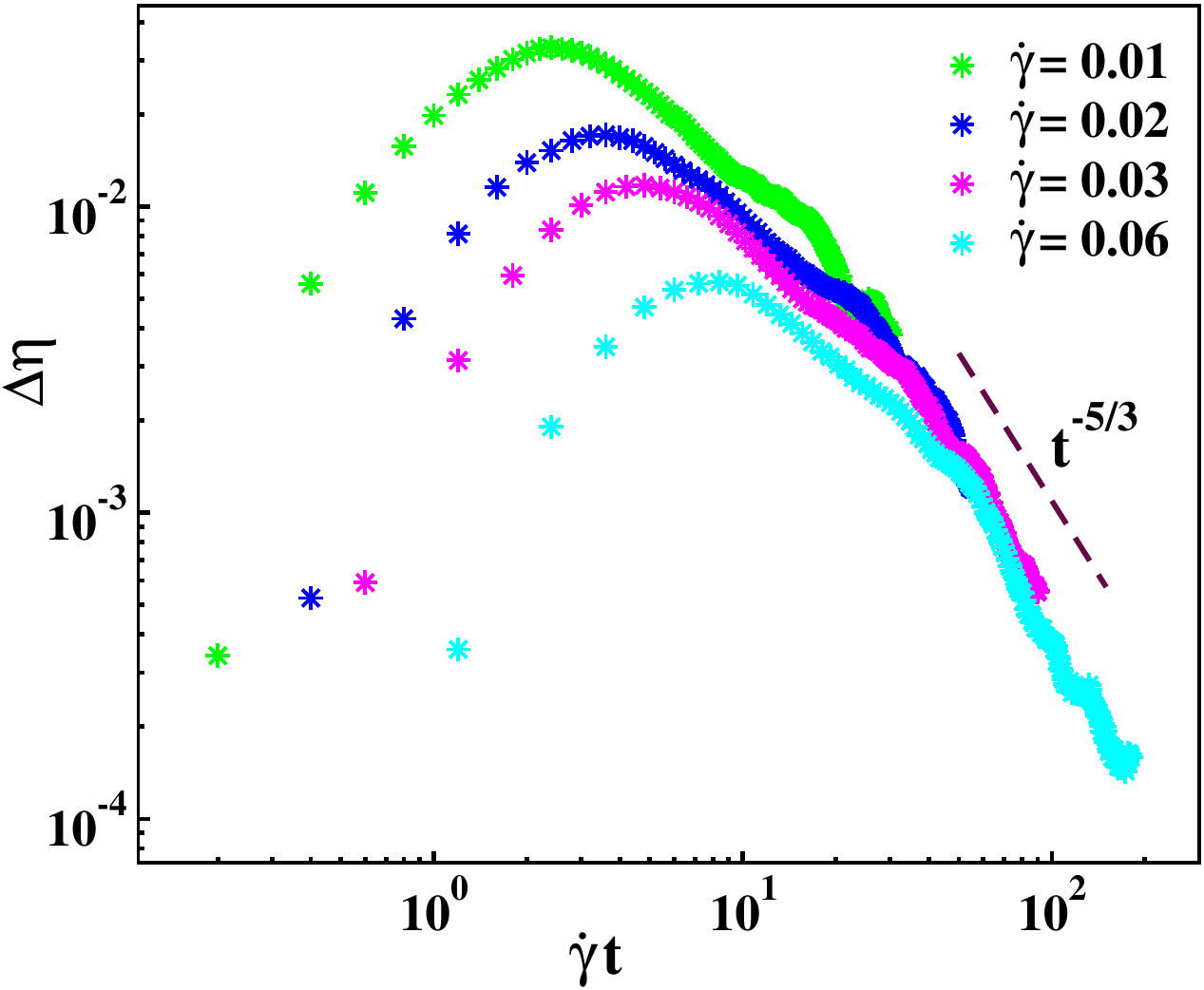}}
    \caption{The excess viscosity $\Delta \eta$ vs shear strain $\dot{\gamma} t$ plot for different shear rates mentioned in the inset.}
    \label{fig:13_excessv_strain}
\end{figure}

The change in shear viscosity of the fluid with strain rate clearly indicates a transition from the Newtonian to non-Newtonian behavior of the system. The domain coarsening contributes to this behavior. To characterize this transition we resort to the excess viscosity expressed as, 
\begin{equation}
	\Delta \eta = \eta_{mix} - \eta_{one}
\end{equation}
We compute the excess viscosity from the structure factor given as
\begin{equation}
	\label{eq:visc1}
	\Delta\eta=-\frac{1}{\dot{\gamma}}\int_{|\vec{k}|<q}\frac{\vec{dk}}{{2 \pi}^d} k_xk_z C(\vec{k},t)
\end{equation}

The variation of $\Delta \eta$ with strain $\dot{\gamma}t$ is shown in Fig.~\ref{fig:13_excessv_strain} for the shear rates $\dot{\gamma} \geq 0.01$. Initially, all the curves increase and reach a maximum value at $t=t_m$ . This is because the domain elongation due to shear involves work against surface tension. The peak position of the curves shift towards lower value of $t_m$ with increasing shear rate. This is attributed to the stronger shear field that aligns the domain interfaces more rapidly in the flow direction. When the excess viscosity plot is compared to the length scale (fig:~\ref{fig:10_lengthscale_0.01}), we observe the asymptotic scaling regime starts when the excess viscosity reaches the maximum at $t=t_m$. This behavior is consistently observed at all high shear rates and agrees satisfactorily with experimental findings \cite{Lauger}.
\begin{figure}[t!]
\centering
   {\includegraphics[width=0.99\columnwidth]{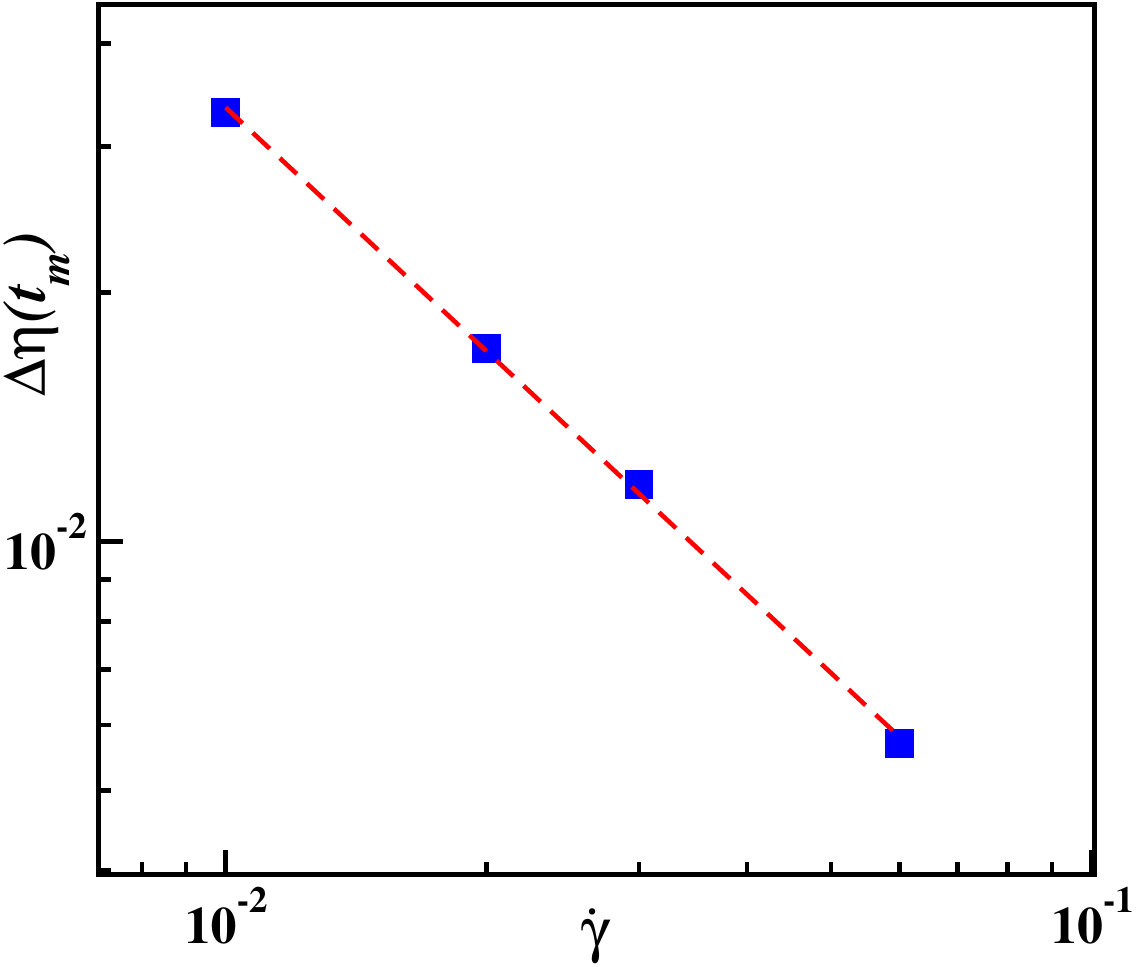}}
    \caption{The plot of excess viscosity $\Delta \eta$ at $t_m$ vs $\dot{\gamma}$. The dashed line has a slope -1.}
    \label{fig:14_shear_thinning}
\end{figure}

At longer time $\Delta \eta$ decreases following a power law $t^{-5/3}$. There are noticeable oscillations in the relaxing region of all the graphs. This can be attributed to growth mechanism of the domains where stretching and breakup of events occur cyclically. The elongation of the domains in the flow direction leads to an increase of $\Delta \eta$. However, as time progresses, the domains become so deformed that they begin to break apart, releasing the stored energy. Consequently, $\Delta \eta$ decreases.

From the discussion so far, it is evident that the maximum of the excess viscosity is a critical point in analyzing both the kinetic and rheological properties of a phase-separating liquid system under shear. Therefore, we examine $\Delta \eta(t_m)$ as a function of the shear rate $\dot{\gamma}$. As shown in Fig.~\ref{fig:14_shear_thinning}, $\Delta \eta(t_m)$ decreases with increasing shear rate, illustrating the phenomenon of shear thinning. The fact that viscosity depends on the shear rate indicates that the liquid exhibits non-Newtonian behavior. We observe a linear relationship for shear thinning in this three-dimensional system.

\section{Conclusion}
In conclusion, we have studied the spinodal decomposition of a three dimensional binary liquid system under simple shear deformation using extensive molecular dynamics simulations. Shear was applied via two moving walls in opposite direction, follow the standard experimental method. A wide range of shear field strength was explored. For the weak shear strength the domain grows isotropically, resembling the no shear case. An increase in shear rate introduced anisotropy into the system, as the domains elongated in the shear direction. The self-similar nature of the domain growth is maintained in this scenario as manifested by the scaling of the spatial correlation functions. The anisotropy was characterized in terms of length scale in the gradient and shear direction. An accelerated domain growth is observed in the shear direction while it is substantially restrained in the gradient direction. The relation in the growth exponents in these two directions is consistent with the theoretical predictions and experimental findings. The growth in the vorticity direction remained oblivious to the shear field. 

We have also investigated the rheological properties of the segregating system in terms of shear stress and excess viscosity. The stress overshoot demonstrated the viscoelastic nature of the system, assisted by domain formation. A transition from a Newtonian to an non-Newtonian behavior was displayed by the system. This was characterized by computing the excess viscosity, which is found to decrease with increasing shear rate. This shear thinning behavior was observed in the high shear field regime. The maximum excess viscosity is related to the threshold of the asymptotic scaling regime, consistent with experimental observations. At long time, the stretching and breakup of domains are manifested as periodic oscillations in the decaying part of the excess viscosity. It will be interesting to study these phenomena under cyclic shear. Studies along this line will be presented elsewhere.

\section*{Acknowledgments} 
B. Sen Gupta acknowledges Science and Engineering Research Board (SERB), Department of Science and Technology (DST), Government of India (no. CRG/2022/009343) for financial support. D. Davis acknowledges VIT for doctoral fellowship. Parameshwaran A. acknowledges SERB India for doctoral fellowship.



\begin{thebibliography}{99}
\bibitem{Binder}
	K. Binder, Phys.Rev.B \textbf{15}, 4425 (1977).
	\bibitem{Siggia}
	E.D. Siggia, Phys.Rev. A \textbf{20}, 595 (1979).
	\bibitem{Furukawa}
	H. Furukawa, Phys.Rev.A \textbf{31}, 1103 (1985).
	\bibitem{Miguel}
	M.San Miguel, M.Grant, and J.D.Gunton, Phys.Rev.A \textbf{32},1001 (1985).
	\bibitem{Tanaka}
	 H. Tanaka, J. Chem. Phys. \textbf{103}, 2361 (1995).
	\bibitem{Beysens}
	  D. Beysens, Y. Garrabos, and D. Chatain, Europhys. Lett. \textbf{86}, 16003 (2009).
	 \bibitem{Tanaka1}
	 S. Tanaka, Y. Kubo, Y. Yokoyama, A. Toda, K. Taguchi,and H. Kajioka, J. Chem. Phys. \textbf{135}, 234503 (2011).
	 \bibitem{Kendon}
	 V.M. Kendon, M.E. Cates, I. Pagonabarraga, J.C. Desplat, and P. Blandon, J. Fluid Mech. \textbf{440}, 147 (2001).
	 \bibitem{Puri}
	 S. Puri, and  B. Dünweg, Phys. Rev. A \textbf{45}, R6977 (1992).
	  \bibitem{Dutt}
	 C. Datt, S.P. Thampi,and R. Govindarajan, Phys. Rev. E \textbf{91}, 010101(R) (2015).
	 \bibitem{Laradji}
	 M. Laradji, and S. Toxvaerd, O.G. Mountain, Phys. Rev. Lett. \textbf{77}, 2253 (1996).
	 \bibitem{Thakre}
	 A.K. Thakre, W.K. den Ohe, and W.J. Briels, Phys. Rev. E \textbf{77}, 011503 (2008).
	 \bibitem{Ahmad}
	 S. Ahmad, S.K. Das, and S. Puri, Phys. Rev. E \textbf{82},  040107 (2010).
	 \bibitem{Min}
	 K.Y. Min, W.I. Goldburg, Phys. Rev. Lett. \textbf{70}, 469 (1993).
	 \bibitem{Hashimoto}
	 T. Hashimoto, K. Matsuzuka, E.Moses, and A. Onuki, Phys. Rev. Lett. \textbf{74}, 126 (1995).
	 \bibitem{Onuki}
	 A.Onuki, J.Phys.:Condens. Matter \textbf{9}, 6119 (1997).
	 \bibitem{Bray}
	 A.J. Bray, Adv. Phys. \textbf{51}, 481 (2002).
	 \bibitem{Chen}
	 C. Chen, W. Liu, H. Wang, and L. Zhu, RSC Adv. \textbf{6}, 102997 (2016).
	 \bibitem{Perot}
	D. Beysens and F. Perot, J. Phys. Lett \textbf{45},  31 (1983).
	\bibitem{chan}
	C. K. Chan, F. Perrot, and D. Beysens, Phys. Rev. Lett. \textbf{61}, 412 (1988).
	\bibitem{Morris}
	D. J. Evans and G. P. Morriss, Phys. Rev. Lett. \textbf{56}, 2172 (1986).
	\bibitem{Onuki1}
A. Onuki, Phys. Rev. A \textbf{35}, 5149 (1987).
	\bibitem{Padilla2}
	P. Padilla and S. Toxvaerd, J. Chem. Phys. \textbf{104}, 5956 (1996).
	\bibitem{Padilla3}
	P. Padilla and S. Toxvaerd, J. Chem. Phys. \textbf{106}, 2342 (1997).
	\bibitem{Corberi1}
	F. Corberi, G. Gonella and A. Lamura, Phys. Rev. Lett. \textbf{81}, 3852 (1998).
	\bibitem{Corberi2}
	F. Corberi, G. Gonella and A. Lamura, Phys. Rev. E \textbf{62}, 8064 (2000).
	\bibitem{Yamamoto}
	R. Yamamoto and X.C. Zeng, Phy. Rev. E \textbf{59}, 3223 (1999).
	\bibitem{Berthier}
	L. Berthier, Phys. Rev. E \textbf{63}, 051503 (2001).
	\bibitem{Cates1}
	P. Stansell, K. Stratford, J.-C. Desplat, R. Adhikari, and M.E. Cates, Phys. Rev. Lett. \textbf{96}, 085701 (2006).
	\bibitem{Cates2}
	K. Stratford, J.-C. Desplat, P. Stansell,  and M.E. Cates, Phys. Rev. E \textbf{76}, 030501 (2007).
	\bibitem{DasTc}
	S. K. Das, M. E. Fisher, J. V. Sengers, J. Horbach, and K. Binder, Phys. Rev. Lett. 97, 025702 (2006).
	\bibitem{Nose}
D. Frenkel and B. Smit, Understanding Molecular Simulations: From Algorithms to Applications (Academic Press, San Diego,2002).  
	\bibitem{Verlet}
      	L. Verlet, Phys.Rev. \textbf{159}, 98 (1967).
	\bibitem{Rounak1}
	R. Bhattacharya and B.S. Gupta, Phys. Rev. E \textbf{104}, 054612 (2021).
	\bibitem{Rounak2}
	R. Bhattacharya and B.S. Gupta, Europhys. Lett. \textbf{140}, 47002 (2022).
	\bibitem{Rounak3}
	R. Bhattacharya and B.S. Gupta, Soft Matter \textbf{20}, 2969 (2024).
	\bibitem{Davis1}
	D. Davis and B.S. Gupta, Phys. Rev. E \textbf{108}, 064607 (2023).
	\bibitem{Davis2}
	D. Davis and B.S. Gupta, arXiv:2308.10295 (2024).
	\bibitem{Paramesh}
	Parameshwaran A and B.S. Gupta, arXiv:2408.01195 (2024).
	\bibitem{gaurav}
	G. P. Shrivastav, S. Krishnamoorthy, V. Banerjee and S. Puri, Europhys. Lett. \textbf{96}, 36003 (2011).
	\bibitem{Majumdar}
	S. K. Das, S. Roy, S. Majumdar, and S. Ahmad, Europhys. Lett. \textbf{97}, 66006 (2012).
	\bibitem{Lauger}
	J. Lauger, C. Laubner and W. Gronski, Phys. Rev. Lett. \textbf{75}, 3576 (1995).
	\bibitem{Corberi3}
	F. Corberi, G. Gonnella, and A. Lamura, Phys. Rev. Lett. \textbf{83}, 4057 (1999).
	\bibitem{Rapapa}
	N. P. Rapapa and A. J. Bray, Phys. Rev. Lett. \textbf{83}, 3856 (1999).
	\bibitem{Bray1}
	A. J. Bray and A. Cavagna, J. Phys. A \textbf{33}, L305 (2000).
	\bibitem{chan1}
	C. K. Chan, F. Perrot, and D. Beysens, Phys. Rev. A \textbf{43}, 1826 (1991).

	
	
\end{thebibliography}
\end{document}